\documentstyle[11pt]{article}

\newcommand{\new}{\newcommand}
\new{\be}{ \begin{equation}}
\new{\ee}{ \end{equation}}
\new{\bea}{ \begin{eqnarray}}
\new{\eea}{ \end{eqnarray}}
\new{\del}{\nabla}
\new{\Del}{\nabla}
\new{\bra}{\left< }
\new{\ket}{\right> }
\new{\E}{{\cal  E}}
\new{\D}{{\cal  D}}
\new{\br}{\bar{r}}
\begin{document}

\begin{flushright}
KAIST-CHEP-96/6
\end{flushright}

\begin{center}
  \Large{ 
 On the Entropy of a Quantum Field in  the 2 + 1 
 Dimensional Spinning Black Holes}
\end{center}
 
\vspace{2cm}
   
\begin{center}
{Min-Ho Lee\footnote{e-mail~:~mhlee@chep6.kaist.ac.kr
}, Hyeong-Chan Kim\footnote{e-mail~:~leo@chep5.kaist.ac.kr},
and Jae Kwan Kim    \\
Department of Physics, Korea Advanced Institute of 
Science and Technology \\
373-1 Kusung-dong, Yusung-ku, Taejon 305-701, Korea.}
\end{center}

\vspace{1cm}

\begin{abstract}
We calculate the entropy of a scalar field in  a
rotating black hole in 2 + 1 dimension. 
In the Hartle-Hawking state  the entropy  is proportional
to the horizon area, but diverges linearly in $\sqrt{h}$, where
$h$ is the radial cut-off. In WKB approximation the superradiant
modes do not contribute to the  entropy.
\end{abstract}

\newpage

 Recently, many efforts have been concentrated on 
understanding the statistical origin of the 
Bekenstein-Hawking black hole entropy \cite{Bek}:
the brick wall method of 't Hooft \cite{tHooft}, the 
entanglement entropy \cite{Ent},
the conical method \cite{conical}, etc.
(See the review \cite{Bekenstein2}.)
The common property of the above methods is that the entropy
is divergent and proportional to the horizon area.

For a rotating black hole in $4$ dimensional space-time
the entropy of a quantum field was calculated by the brick 
wall method \cite{minho}. The result is that 
the entropy is proportional to the horizon area 
in the Hartle-Hawking state. 
The difficulty in  treating the quantum field in a rotating 
black hole  background  is that one can not find a  global 
static frame.  Usually one resolve it by taking a rigid
frame co-rotating  with the  black hole.
However in this case an observer who is at the outside of a surface 
(the velocity of light  surface (VLS) ) must 
have $v \geq 1$ and must  move on a spacelike world line.
To remove such an unphysical behavior one needs a perfectly
reflecting mirror inside  the VLS \cite{Thorne}.

In $3$ dimension Banados, Teitelboim, and Zanelli (BTZ)
  obtained a black hole
solution for the standard $2+1$ Einstein-Maxwell theory with 
a negative cosmological constant, which (for charge = 0) is 
asymptotically
anti-de Sitter space-time \cite{banados}. This is also
the  solution of  the low energy string action in 3
dimension \cite{horowitz}.
Chan and Mann modified the BTZ black hole and obtained a new class of
spinning black hole solutions \cite{chan}.
The black hole is characterized
by mass, angular momentum, and charge, which is similar to the $4$ 
dimensional rotating black hole.
Therefore to study the 3 dimensional black hole is helpful
to understand the entropy of the 4 dimensional black hole.

The entropy of the BTZ black hole was calculated by the 
brick wall method in Ref. \cite{wtkim}. They found that
the entropy is finite for the BTZ black hole with non-zero
angular momentum.
In this paper we study the entropy of a quantum field in
3 dimensional spinning black hole \cite{chan}
 by the brick wall method.
We  show that the entropy  diverges linearly in $\sqrt{h}$, where
$h$ is the radial coordinate distance from the horizon to the 
brick wall. 
In WKB approximation the  superradiant modes 
in the Hartle-Hawking state do not contribute to the 
 entropy.


Let us consider  a scalar field with mass $\mu$
in  thermal equilibrium at  temperature $1/\beta$  in a
 rotating 3 dimensional black hole background, 
 of which line element is generally given by
\be
ds^2 =  g_{tt}(r) d t^2  
    + 2 g_{t \phi}( r) dt d \phi 
    +  g_{\phi \phi}(r) d\phi^2 
    + g_{rr} (r) d r^2. 
               \label{metric}  \\
\ee 
This metric has two Killing vector fields: the timelike 
Killing vector $\xi^\mu = (\partial_t)^\mu$ and the axial
Killing vector $\psi^\mu =(\partial_\phi)^\mu$.
In this paper we consider the spinning  black hole with the following
metric components \cite{chan}
\bea
\nonumber
g_{tt} & =& - \left( 
       \frac{8 \Lambda \alpha^2 }{(3N -2)N} r^N + A r^{1 - \frac{N}{2}}
       \right), \\
g_{t \phi} &=& - \frac{\omega}{2} r^{1 - \frac{N}{2}}, 
 \label{metric2} \\ 
\nonumber
g_{\phi \phi} &=& 
        \left(  \alpha^2 r^N - \frac{\omega^2}{4 A} r^{1 - \frac{N}{2}}
	\right), \\
\nonumber
g_{rr} &=& \alpha^2 \left[
        \frac{8 \Lambda \alpha^2}{(3N-2)N } r^N + \left(
	A - \frac{2 \Lambda \omega^2}{(3N-2)N A} \right)
	r^{1 - \frac{N}{2}}
	\right]^{-1},
\eea
where $A$ and $\omega$ are integration constants and
$\alpha$ is a length scale with dimensions of length.
The mass and the angular momentum of the 
black hole is given by  
\begin{eqnarray}
M &=& \frac{N}{2} \left[
\frac{2 \Lambda \omega^2}{(3N-2)N A} \left(
\frac{4}{N} -3 \right) - A \right], \\
J &=& \frac{3N-2}{4} \omega.
\end{eqnarray}
The black hole exist if $\Lambda >0$ and $2 \geq N > \frac{2}{3}$.
The constant $A$  is negative so that $g_{\phi \phi} >0$.
For  this spinning  black  hole  there are two important 
surfaces: the outer   horizon and the
stationary limit surface.
The outer  horizons are given  by 
\be
r_+^{\frac{3N}{2} - 1} =   \left(
	 \frac{8 \Lambda \alpha^2}{(3N-2)N }
	\right)^{-1} 
	\left[ - A + \frac{2 \Lambda \omega^2}{(3N-2)N A }
	\right],
\ee
and the stationary  limit surface is given by
\be
r_s^{\frac{3N}{2} -1} =
	 \left( \frac{8 \Lambda \alpha^2}{(3N-2)N }
	 \right)^{-1}  (-A).
\ee
The Killing vector $\xi^\mu$ vanishes  on the stationary limit surface, 
and the Killing vector $\xi^\mu + \Omega_H \psi^\mu$ is null  on
the event horizon $(r = r_+ = r_H)$, where $\Omega_H$ is the angular 
velocity of the horizon \cite{wald}:
\be
\Omega_H = \lim_{r \rightarrow r_H}  \left(
  - \frac{g_{t \phi}}{g_{\phi \phi}} \right) 
  = - \frac{4 \Lambda \omega}{(3N- 2)N  A }.
\ee
When $N =2$, this black hole solution reduces to the  spinning BTZ 
one.
When $N= 1$,  this solution becomes the modification of  the black 
hole of Mandal, Sengupta, and Wadia~\cite{Mandal}. 

The equation of motion of   the  field  with mass $\mu$
is given by 
\be
\left[ \Del_\mu   \Del^\mu   -  \xi R - \mu^2 \right] 
\Psi = 0,           \label{equation}
\ee
where  $\xi$ is an arbitrary  constant
and $R(x)$ is  the scalar curvature. 
$\xi = 1/8$ and $\mu =0$ case corresponds to the conformally
coupled one.
We assume that the scalar field  is rotating with a constant 
azimuthal angular velocity $\Omega_0 \leq \Omega_H$. 
The  associated  conserved quantity is angular momentum. 
The positive frequency field mode can be written as 
$\Phi_{q,m} = f_{q,m}(r) e^{- i \E t + i m \phi }$,
where $m$ is the azimuthal quantum number and 
$q$ denotes other  quantum numbers. 
The free energy  of the system is then given by 
\be
 F  =  \frac{1}{\beta} \sum_{j,m} d_{j,m} \ln 
   \left( 1 - e^{- \beta( \E_{j,m} - m \Omega_0  )}  
   \right)  \label{po}
\ee
or
\be
   F =   \frac{1}{\beta} \sum_m \int_0^\infty d \E  g(\E, m) \ln 
       \left( 1 - e^{- \beta( \E - m \Omega_0  )}   \right), 
\ee
where $g(\E,m)$ is the density of state for a given $\E$ and $m$.

To evaluate the free energy we  follow 
the brick wall method  of 't Hooft \cite{tHooft}.
We impose a small radial cut-off $h$ such that 
\begin{equation}
\Psi  (x) = 0 ~~~~{\rm for  }~~~ r \leq r_H + h,
\end{equation}
where $r_H$  denotes the coordinate of the event horizon.
To remove the infra-red divergence   we also introduce another 
cut-off  $ L \gg r_H$ such that 
\be
\Psi (x) = 0~~~~ {\rm for} ~~~r \geq L.
\ee
In the WKB approximation with $\Psi = 
e^{- i \E t + i m \phi + i S(r) }$
the  equation (\ref{equation})  yields 
the constraint \cite{Mann}
\begin{equation}
  p_r^2 = \frac{1}{g^{rr}}  \left[
       - g^{tt} \E^2 + 
       2 g^{t \phi} \E  m - g^{ \phi \phi } m^2 
       - V(x)      \right],         \label{Con1}
\end{equation}
where  $  p_r = \partial_r S$ and
$V(x) = \xi R(x) + \mu^2$.
It is important to note that the number of state for a 
given  $\E$ is determined by  $ p_r$ and $m$.
The number of mode with energy less than $\E$ and  with a fixed
$m$ is obtained by integrating over the  phase space 
\bea
\nonumber
\Gamma (\E,m )  &=& \frac{1}{\pi} \int d \phi  \int dr
 p_r ( \E, m,x)   \\
&=& \frac{1}{\pi} \int d \phi  \int dr
  \left[ \frac{1}{g^{rr}}  \left(
       - g^{tt} \E^2 + 
       2 g^{t \phi} \E  m - g^{ \phi \phi } m^2 
       - V(x)      \right) 
       \right]^{\frac{1}{2}}.        
\eea

At this point we need some remarks.
In a rotating system, in general, there is a  superradiance effect,
which occurs when $ 0 < \E < m \Omega_0$.
For this range of the frequency the free energy $F$ becomes a complex
number. In case $\E = m \Omega_0$ the free energy is divergent.
Therefore  to obtain a real finite value for the free energy $F$,
we must require that $\E > m \Omega_0$. ( For $ 0 < \E < m \Omega_0$
the free energy diverges. See below.)  This requirement says that
we must restrict the system to be in the region 
such that $g_{tt}^{'} \equiv g_{tt} + 2 \Omega_0 g_{t \phi} 
+ \Omega_0^2 g_{\phi \phi} < 0$. 
In this region 
the free energy is a finite  real  value
because $ \E - m \Omega_0 > 0$.
It is easily  shown  as follows.
Let us define $ E = \E - m \Omega_0$.
Then it is written as 
\bea
\nonumber
E &=& \left(
         \frac{g^{t \phi}}{ g^{tt}  } -  \Omega_0 
      \right) m
        + \frac{1}{- g^{tt}} 
     \left[
       \left( 
         g^{t \phi} m 
       \right)^2 
              + \left( - g^{tt}    \right)
       \left(    V + g^{\phi \phi} m^2 + 
          g^{rr} p_r^2   
       \right)
     \right]^{1/2} \\
  &=& \left( 
         \Omega - \Omega_0 
     \right)  m   + \frac{ -\D }{g_{\phi \phi}}
     \left[
         \frac{1}{-\D} m^2   +   \frac{ g_{\phi \phi} }{- \D }
       \left( V + \frac{p_r^2 }{g_{rr}}
       \right)
    \right]^{1/2},   \label{Con2}
\eea
where we used 
\be
g^{tt} = \frac{g_{\phi \phi}}{ \D},~~
g^{t \phi} = \frac{ - g_{t \phi}}{\D},~~
g^{\phi \phi} = \frac{ g_{tt}}{\D},
\ee
and  $ \Omega = -\frac{ g_{t \phi}}{g_{\phi \phi}} $.
Here $ - \D = g_{t \phi}^2 - g_{tt}g_{\phi \phi}$.
From Eq.(\ref{Con2}),  for all $m$ and  $p_r$   
one can see that the condition such that $ E >0$ is
\be
 \frac{  \sqrt{-\D }  }{ g_{\phi \phi}}  \pm 
\left( \Omega  - \Omega_0    \right) > 0
\ee
or
\be
g_{tt}^{'} \equiv g_{tt} + 2 \Omega_0 g_{t \phi} 
+ \Omega_0^2 g_{\phi \phi} < 0.
\ee
 Therefore in the region such that $ - g_{tt}^{'} >0$ 
 ( called region I)  the free energy is  real, but
 in the region such that $- g_{tt}^{'} < 0$ (called region II)
  the  free energy is complex.
However in the region  II the integration over the momentum 
phase space is divergent.
 This fact  becomes  apparent if we investigate  the momentum 
 phase space.
  In the region  I 
the  points of $p_i$  satisfying $ \E  - \Omega_0 p_\phi  = E$ 
for a given $E$ are located on the following curve
\begin{equation}
\frac{p_r^2}{g_{rr}} + 
      \frac{- {g'}_{tt}}{- \cal D} \left(
              p_\phi + \frac{g_{t \phi }  
           + \Omega_0 g_{\phi \phi}}{{g'}_{tt}} 
           E  \right)^2 
   = \left( \frac{ E^2}{- {g'}_{tt} } 
   - V \right),   \label{ellipsoid}
\end{equation}
which is the ellipse,  {\it a closed curve}. 
Here $p_\phi = m$.
So the density of state $g(E)$ for a given $E$ is finite and 
the integrations over $p_i$  give a  finite value.
But in the region  II 
the points of $p_i$  are located on the following curve
\begin{equation}
\frac{p_r^2}{g_{rr}} - 
        \frac{ {g'}_{tt}}{- \cal D} \left( 
         p_\phi + \frac{g_{t \phi }  + 
         \Omega_0 g_{\phi \phi}}{{g'}_{tt}} 
	 E     \right)^2 
   =  - \left( 
             \frac{ E^2  }{  {g'}_{tt}} 
             + V \right),
\end{equation}
which is  the hyperbola,   {\it a open curve}. So 
$g(E)$ diverges and the integrations over $p_i$  diverge. 
In case of $g^{'}_{tt} = 0$, the points of $p_i$  are 
given by 
\begin{equation}
\frac{p_r^2}{g_{rr}} = 
\frac{p_\phi - \left( \frac{ g_{\phi \phi} E^2 }{ \cal D }  + V 
            \right)/ \left( \frac{ 2 g_{t \phi} }{ \cal D } E 
	            \right) 
      }{ \frac{- {\cal D} }{ 2 g_{t \phi} E }  
       },
\end{equation}
which is a parabola and also $open~ curve$. Therefore 
the value of the $p_i$ integrations are divergent.
Actually the surface (the curve) such that ${g'}_{tt} = 0$ 
is the velocity of the
light surface (VLS). Beyond VLS (in region II) the co-moving 
observer must move  more rapidly  than the
velocity of light. It is unphysical.
Thus we   assume  that the  system is in the region I.

Now let us determine the  region I.
From 
\bea
\nonumber
g'_{tt} &=& g_{tt} + 2 \Omega_0 g_{t \phi } + 
         \Omega_0^2 g_{\phi \phi}   \\
	&= &   - \left( 
	\frac{8 \Lambda \alpha^2}{(3N-2)N}
	-  \Omega_0^2  \alpha^2 \right) r^{N} 
	+ \left(
	- A - \Omega_0 \omega  - \Omega_0^2 \frac{\omega^2}{4 A}
	\right) r^{1 - \frac{N}{2}} 
\eea
we obtain  the exact position of the VLS, which  is given by
\be
  r_{VLS}^{\frac{3N}{2} -1}  = \left[
	\frac{8 \Lambda \alpha^2}{(3N-2)N}
	-  \Omega_0^2  \alpha^2  \right]^{-1}
	\left[
	- A - \Omega_0 \omega  - \Omega_0^2 \frac{\omega^2}{4 A}
	  \right].
\ee
For $\Omega_0 = 0$    the  VLS  is  at $r = r_s$, and 
for $\Omega_0 = \Omega_H$   it locates at $ r = r_+$.
As  the value of $\Omega_0$ increases from $0$ to $\Omega_H$ 
the VLS is  continuously moved from $r_s$ to $r_+$.
But there is no outer VLS,
which is  distinct from the 4-dimensional black hole \cite{minho}.  
Thus the region I is $  r_{VLS} < r < \infty$.


With  the assumption that the system is in the region I 
we   obtain the free energy  as follows:
\begin{eqnarray}
\nonumber
  \beta F  &=& \sum_m \int_{m \Omega_0}^\infty d \E  g(\E, m) \ln 
       \left( 1 - e^{- \beta( \E - m \Omega_0 )}   \right)    \\
\nonumber
       &=& \int_0^\infty d\E \sum_m g(\E + m \Omega_0 , m) \ln 
       \left( 1 - e^{- \beta \E }   \right)             \\
       &=&  - \beta \int_0^\infty d\E \frac{1}{e^{\beta \E} - 1}
            \int d m \Gamma (\E + m \Omega_0, m),
\end{eqnarray}
where we have integrated by parts and we assume that the quantum 
number $m$ is a continuous variable.   
The integration over $m$  yields 
\begin{equation}
F =   -   
\int d \phi  \int_{r_H + h}^L dr
\int_{V(x) \sqrt{- {g'}_{tt}}}^\infty d\E  
\frac{1}{e^{\beta \E} - 1 }
\frac{ \sqrt{g_3}}{\sqrt{ - {g'}_{tt} }}  \left( 
\frac{\E^2}{ -  {g'}_{tt} }  - V(x)  \right).  
\label{freeenergy}
\end{equation} 
In particular when $\Omega_0 = 0$,  $J = 0$, and $V(x) =0$, 
the free energy (\ref{freeenergy}) 
 is proportional to the volume of the optical space \cite{optical}.
It is easy to see that the  integrand diverges  as  
$ r_H + h $  approaches   $r_{VLS}$.  
In that case  the contribution  of the $V(x)$ can be negligible. 

In the case of $V = 0$
the free energy reduces to 
\begin{equation}
\beta F = - \frac{c}{\beta^2 }  \int  d \phi 
\int_{r_H + h}^L dr \frac{\sqrt{g_3}}{ ( - g^{'}_{tt } )^{3/2}  } 
 = - c \int_0^\beta d \tau \int  d \phi 
\int_{r_H + h}^L dr \sqrt{g_3} 
 \frac{1}{ \beta_{local}^3},            \label{freeenergy2}
\end{equation}
where $\beta_{local} = \sqrt{ - g^{'}_{tt } } \beta$ is 
the reciprocal of the local Tolman temperature \cite{Tolman}
in the comoving frame, and $c$ is  a constant  .
This form  is just the free energy of a gas of 
massless particles at local
temperature $1/\beta_{local}$ in 3 dimension.


From this expression (\ref{freeenergy2})  
it is easy to obtain the expression for 
the entropy   $S$ of a scalar field for $V(x)  = 0$. 
In the Hartle-Hawking state ( $\Omega_0 = \Omega_H, 
T = T_H$),  where
\be
T_H = \frac{\Lambda \alpha^2}{\pi N}
\frac{r_H^{\frac{3N}{2} -1} }{
\sqrt{g_{\phi \phi}(r_H) }} = \frac{1}{\beta_H}
\ee
the entropy for small $h$ is  given by 
\begin{eqnarray}
\nonumber
S &= & \left. \beta^2 \frac{\partial}{\partial \beta } F 
    \right|_{\beta = \beta_H, \Omega_0 = \Omega_H} \\
\nonumber
&=& \frac{6 \pi c \alpha^2 }{\beta_H^2}
\int_{r_H + h}^L dr 
\frac{r^\frac{N}{2} }{
\left[
    \left(
	\frac{8 \Lambda \alpha^2}{(3N-2)N}
	-  \Omega_H^2  \alpha^2 
    \right) r^{N} 
    - \left(
	- A - \Omega_H \omega  
	- \Omega_H^2 \frac{\omega^2}{4 A}
      \right)
     r^{1 - \frac{N}{2}  } 
     \right]^{\frac{3}{2} } }
	\\
&\approx&  
  \frac{3 c \pi }{ 2 \alpha^4 \beta_H^2}
  \left( \frac{ N}{\Lambda} \right)^\frac{3}{2}
r_H^{\frac{3}{2} - \frac{5}{2} N} g_{\phi \phi}^{\frac{3}{2}} (r_H)
\frac{1}{ \sqrt{h}} + O(\sqrt{h}).
\label{entropy1}
\end{eqnarray}
The entropy is linearly divergent in $\sqrt{h}$. (This is the
general feature of the non-degenerated 2 + 1 dimensional black hole.)
In terms of  the proper distance cut-off $\epsilon$  the 
entropy is given by  
\be
S = \frac{3 c}{8 \pi^2}  \frac{A_H}{\epsilon},
\label{entropy2}
\ee
where  
\be
A_H = 2 \pi \sqrt{ g_{\phi \phi}(r_H) }
= 2 \pi 
  \left( \alpha^2 r_H^N - \frac{\omega^2}{4A} r_H^{1 - \frac{N}{2} }
\right)^{\frac{1}{2}}
\ee
and
\bea
\nonumber
\epsilon &=& \int_{r_H}^{r_H + h} dr  \sqrt{g_{tt} (r)}\\
&\approx&  \left( 
\frac{ N}{\Lambda r_H^{N-1}} \right)^{\frac{1}{2}} \sqrt{h}. 
\eea
Notice that  the entropy S (\ref{entropy2}) does not 
depends on the  constants
$\alpha, N, A,$ and $ \omega$.  
In 4 dimensional black hole we also  showed that the leading behavior
of the entropy of the 
quantum field is proportional to the horizon area and is not depend on 
the parameters  if we  use the
proper distance cut-off $\epsilon$ \cite{minho}.
It seems that this is a generic feature of the entropy.
In Ref. \cite{wtkim} 
they argued that the reality  condition of  the radial mode 
momentum  is that the brick wall must be outside of 
the stationary limit surface and they said that the entropy is finite.
In this paper, we demand that the free energy is finite.
This condition is satisfied if the brick wall is present  
outside of the horizon. In this case,
the entropy diverges as the cut-off go to zero.

Let us summarize our result.  We have calculated the 
entropy of the scalar field in the 2+1 dimensional spinning
black hole space-time.
For the massless field in the Hartle-Hawking state 
($\Omega_0 =\Omega_H$ and $ \beta = \beta_H$), 
only in this case, the entropy  
is proportional to the horizon area, but becomes divergent
linearly in $\sqrt{h}$ as 
the brick wall approaches the horizon.  The origin of the 
divergence is  that the momentum phase volume for a given
$E$ diverges on the horizon. 
For the extreme BTZ black hole $(r_+ = r_-)$ (N = 2 case) , 
${g'}_{tt}(r)  |_{\Omega_0 = \Omega_H } =  0  $.
So we can not consider the extreme one. 

Why there is no the outer velocity of light surface in spinning $2 +1$ 
dimensional black hole?  In 4 dimensional black hole the outer VLS 
exists, which  show the pathology  of the rigid rotation of 
the frame.  The 4 dimensional black hole space-time is asymptotically
flat and non-rotating. But  the space-time, for example,  of the 
BTZ black hole is asymptotically anti-de Sitter and have a
rotation.
\be
ds^2_{BTZ} \stackrel{r \rightarrow \infty}{\longrightarrow}
- \left( \frac{r^2}{l^2} - M \right) dt^2 - J dt d \phi
 + \frac{1}{ \frac{r^2}{l^2} - M }dr^2   +  r^2 d \phi^2.
\ee
The $g_{t \phi}$ does not vanish at the infinity.
It is a constant.
Such a fact seems to be  a  cause of the 
non-existence of the outer VLS.


\begin{flushleft}
{\bf Acknowledgment}
\end{flushleft}

This work is partially supported by Korea Science and Engineering 
Foundation.

\end{document}